\documentstyle[sprocl]{article}

\input{psfig.sty}

\begin{document}

\title{The Impact of Tidal Interactions on Satellite Galaxies: A Study
of the M31 Satellites, M32 \& NGC~205}

\author{P. I. Choi, P. Guhathakurta} 
\address{UCO/Lick Observatory, University of California, Santa Cruz, CA 95064\\
E-mail: pchoi@ucolick.org, raja@ucolick.org}

\author{K. V. Johnston}
\address{Van Vleck Observatory, Wesleyan University, Middletown, CT 06459\\
E-mail: kvj@astro.wesleyan.edu}

\maketitle\abstracts{Surface photometry of the M31 satellites M32 and
NGC~205 is compared to numerical simulations of satellite destruction
to constrain orbital parameters and the interaction history of the M31
subgroup.  Our analysis reveals the following preliminary results: (1)
Generic features of tidal disruption in the simulations include an
extended ``extra-tidal'' excess region and an inner depletion zone,
both of which are observed in M32 and NGC~205; (2) M32 is likely to be
on a highly eccentric orbit well away from pericenter; (3) Surface
brightness and luminosity evolution estimates for M32, the
prototypical compact elliptical galaxy, imply that it is not simply
the residual core of a tidally-stripped normal elliptical galaxy, but
was instead formed in a truncated state.}

\section{Introduction}
Recent evidence for tidal streams in the halos of the Milky
Way~\cite{leon+00} and M31~\cite{ibata+01}, along with studies
investigating extra-tidal material around Local Group globular
clusters~\cite{g+95} and dwarf spheroidals~\cite{majewski+00},
indicate that the tidal disruption and accretion of satellites are
ongoing processes in the present epoch.  In this paper, preliminary
results are presented from a study comparing integrated surface
photometry of M32 and NGC~205 to satellite
simulations~\cite{paper2a}$^{,}$~\cite{paper2b}.  Spectroscopic
observations to determine the internal kinematics in the tidal region
of M32 and more finely tuned numerical models will be incorporated in
later studies.

\section{Observations / Simulations}

\begin{figure}[t]
\vskip 2.5cm
\vskip -3.1cm
\psfig{figure=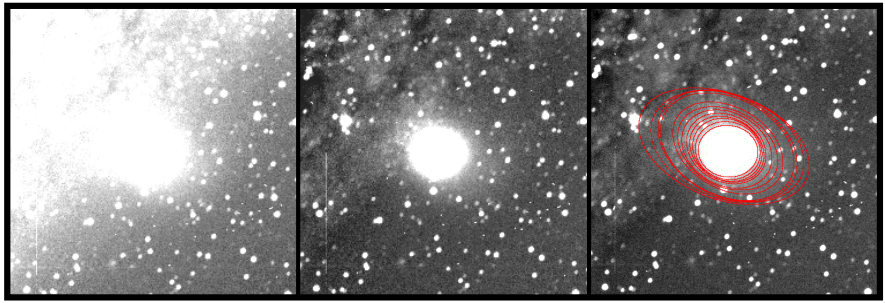,height=1.6in,angle=0}
\caption{Grayscale representations of $B$-band images
($17^{\prime}\times 17^{\prime}$) centered on M32 with ({\it left})
and without ({\it middle}) M31's disk light contribution.  Note the
steep gradient in the background across M32 caused by the inclined
disk of M31 ({\it left}) and M31's residual fine-scale structure [dust
lanes, spiral arms, etc.~({\it middle})] even after subtraction.
Best-fit elliptical isophotes of M32 ({\it right}) in the semi-major
axis range of $100^{\prime\prime}<r<300^{\prime\prime}$ highlight the
low surface brightness region in which signatures of tidal
interactions are observed.
\label{fig1}}
\vskip -0.25cm
\end{figure}

The observational component of this study is based on
$1.7^{\circ}\times5^{\circ}$ $B$- and $I$-band CCD mosaic images
centered on M31 and covering both satellites.  Standard ellipse
fitting techniques are used to model and remove M31 disk light and to
perform surface photometry on the satellites to limiting isophotes of
$[\mu_B,~\mu_I]=[27,~25]~{\rm mag~arcsec}^{-2}$, corresponding to
semi-major axis lengths $r_{\rm lim}^{\rm M32}=420^{\prime\prime}$
(1.6~kpc) and $r_{\rm lim}^{\rm NGC~205}=720^{\prime\prime}$
(2.7~kpc)~\cite{paper2a}. In Figure \ref{fig1}, $B$-band images of
M32, before and after M31 subtraction, illustrate the importance of
careful background subtraction.  For the numerical simulations,
single-component, spherical satellites are followed through a fixed
three-component potential representative of the disk, bulge and halo
of a parent galaxy.  The models explore a range of orbital
eccentricities and initial mass profiles for the satellite.  To
facilitate comparison, the ellipse fitting technique used for the
observations is also adopted for the simulated satellites.

\begin{figure}[t]
\vskip 2.5cm
\vskip -3.1cm
\psfig{figure=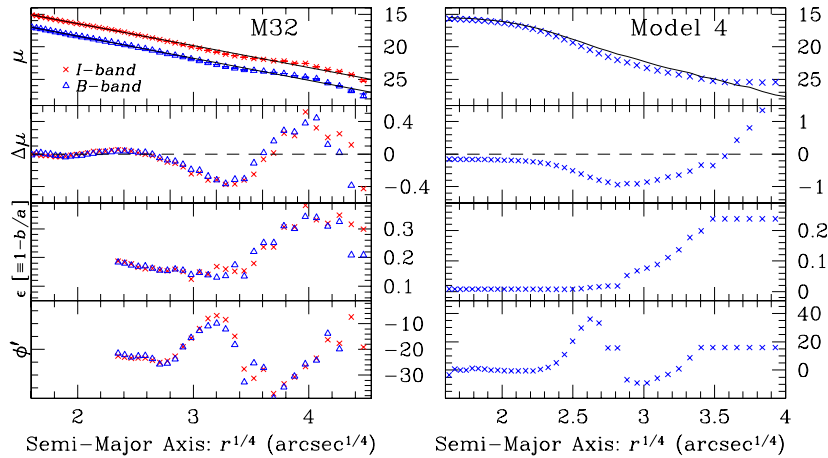,height=2.55in}
\vskip -0.0cm
\caption{Surface brightness ${\mu}$, de~Vaucouleurs residual
${\Delta\mu}$, ellipticity ${\epsilon}$ and position angle ${\phi}$
profiles plotted in de~Vaucouleurs coordinates for a simulation
snapshot ({\it right}) and M32 ({\it left}) in $B$ ({\it triangles})
and $I$ ({\it crosses}) bands.  The surface brightness profile of M32
is well fit by a de~Vaucouleurs profile ({\it top left}) over the
radius range $10^{\prime\prime}<r<65^{\prime\prime}$ with $r_I^{\rm
eff} \sim r_B^{\rm eff}\sim30^{\prime\prime}$ and [$\mu_I^{\rm
eff},\mu_B^{\rm eff}$]=[$18.0,19.9$]~mag~arcsec$^{-2}$.  The residual
profile shows a deficiency below the extrapolated de~Vaucouleurs fit
from $50^{\prime\prime}-150^{\prime\prime}$ and an excess beyond
$150^{\prime\prime}$.  These features, coincident with breaks in the
ellipticity and position angle profiles, are comparable to those found
in Model~4 ({\it right}) in which a satellite with orbital
eccentricity $e=0.88$ is approaching apocenter.  The bold line
covering the range $100^{\prime\prime}<r<300^{\prime\prime}$ in the
M32 $\mu$ plot ({\it top left}) shows the region marked by contours in
Figure~1 (\it right).
\label{fig2}}
\vskip -0.25cm
\end{figure}

\begin{figure}[t]
\vskip 2.5cm
\vskip -3.1cm
\psfig{figure=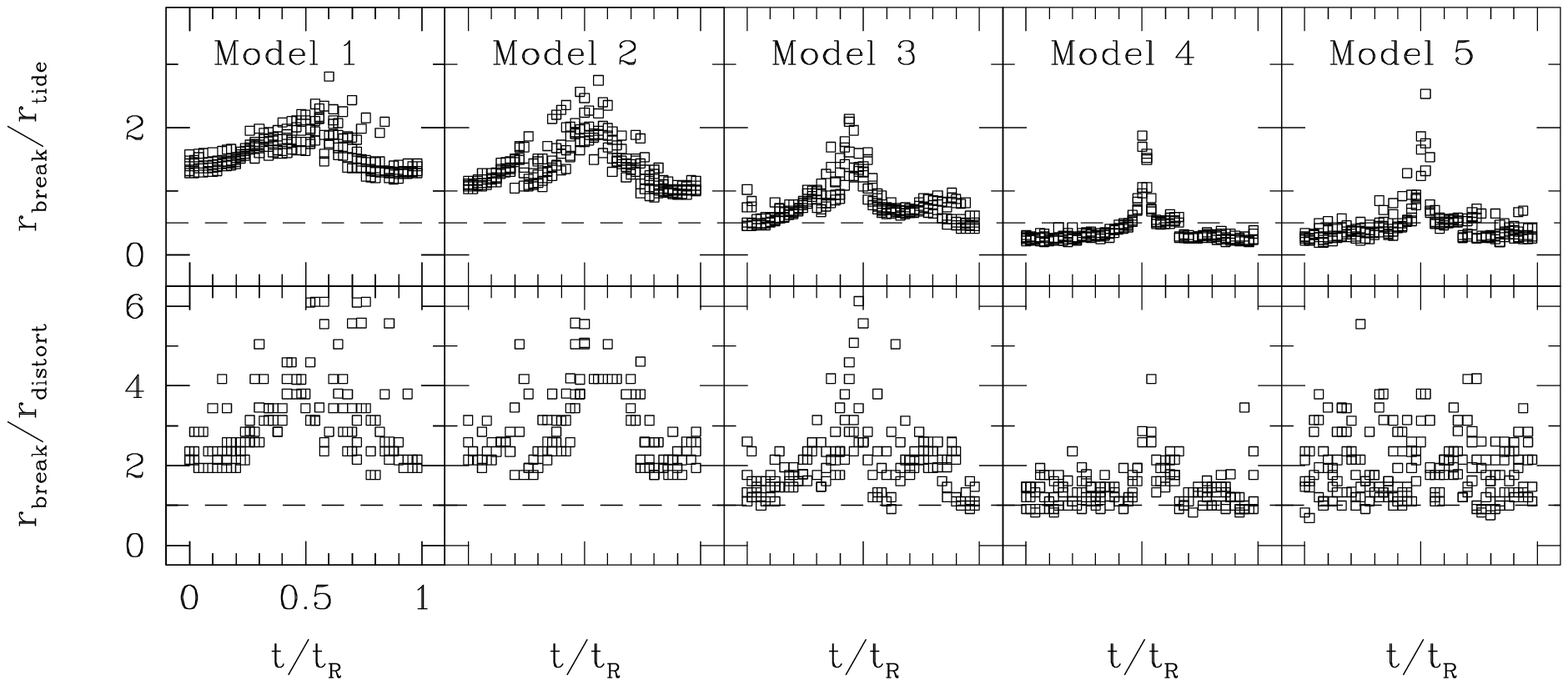,height=2.07in}
\vskip -0.3cm
\caption{Ratio of $r_{\rm break}/r_{\rm tidal}$ ({\it top}) and
$r_{\rm break}/r_{\rm distort}$ ({\it bottom}) as a function of
orbital phase.  In Models 1--4, satellite orbits range in eccentricity
from nearly circular (Model~1) to highly elongated (Model~4) with
eccentricities of $e=0.10/0.29/0.67/0.88$, respectively.  Model 5
follows the same orbit as Model~4, but adopts a shallower initial
density profile than Models 1--4.  The dashed line in each panel
represents the measured ratio for M32 and indicates that it is likely
to be on an eccentric orbit ($e_{\rm M32}\ge0.5$).
\label{fig3}}
\vskip -0.3cm
\end{figure}

\section{Interpretation of Observations in Light of Numerical Simulations}
\underline{Generic Features of Tidal Interaction}: Breaks in the
surface brightness, ellipticity and position angle profiles are common
features of the numerical simulations (Fig.~2: {\it right}).  The
presence of an extended region of excess material and an inner
depletion zone are other generic features.  The excess region
corresponds loosely to what is conventionally described as an
``extra-tidal'' region; however, we find that in many cases they are
associated with tidally heated, yet bound material.  Similar features
are observed in M32 (Fig.~2: {\it left}) and NGC~205 suggestive of
tidal interaction with and stripping by M31 \cite{paper2b}.

\vskip 0.2cm
\noindent \underline{Discriminating Orbital Parameters for M32}: In
the case of M32, three of its profile features are suggestive of a
highly eccentric orbit:
\begin{itemize}
\item{There is a triple break in the position angle $\phi$ profile,
with two of the breaks coincident with breaks in the surface
brightness and ellipticity profiles.  This is an atypical feature of
the simulations, seen only in satellites approaching apocenter on
highly eccentric orbits.  In Figure~2, the profiles of one such
simulated satellite show striking similarities to those of M32.}

\item{The second piece of evidence is the relationship between the
classically defined, theoretical King tidal radius $r_{\rm tide}$ and
the observed break in the surface brightness profile $r_{\rm
break}$~\cite{paper2b}.  The ratio $r_{\rm break}/r_{\rm tidal}$
typically has values of unity or greater for near-circular orbits and
only drops below unity for certain phases of highly eccentric orbits
(Fig.~3: {\it top}).  For M32, the measured upper limit of $r_{\rm
break}/r_{\rm tidal}\sim0.5$ suggests that it is on a highly eccentric
orbit away from pericenter.}

\item{Finally, $r_{\rm distort}^{\rm M32}$, the radius of the onset of
isophotal elongation is coincident with $r_{\rm break}^{\rm
M32}\sim150^{\prime\prime}$. This ratio, $r_{\rm break}/r_{\rm
distort}$ is typically $\geq2.0$ for near-circular orbits and
approaches unity only for the most eccentric orbits (Fig.~3: {\it
bottom}), suggesting that M32 is in this latter category. Since both
radii are directly observable, unlike $r_{\rm break}/r_{\rm tidal}$,
this deduction is less model dependent and more robust than the
previous one.}
\end{itemize}

\vskip 0.2cm
\noindent \underline{Implications on Compact Elliptical Galaxy
Formation}: M32 stands apart from normal ellipticals---its combination
of high central surface brightness and low luminosity make it the
prototype of a rare class of galaxies known as compact ellipticals
(cEs).  cEs tend to reside in close proximity to massive companion
galaxies, and it is commonly believed that cEs are the tidally
truncated remnant cores of normal ellipticals.  Guided by numerical
simulations, we estimate the luminosity $L^{\rm orig}$ and central
surface brightness $\mu_0^{\rm orig}$ M32 might have had {\it prior
to\/} tidal stripping by M31.  These estimates are compared to the
observed (i.e.~present-day) values of $L$ and $\mu_0$ for M32 in order
to infer the changes $\Delta L$ and $\Delta\mu_0$.  While the changes
are in the right direction they are far too small to explain M32's
position in the $L$-$\mu_0$ plane.  This suggests that the true impact
of environment may be in the formation, rather than in the subsequent
evolution, of cEs~\cite{burkert93b}.


\section*{References}
\begin{thebibliography}{99}

\bibitem{burkert93b}A. Burkert, MNRAS {\bf 266}, 877 (1993)

\bibitem{paper2a}P.I. Choi, P. Guhathakurta \& K.V. Johnston, AJ submitted (2002)

\bibitem{g+95}C.J. Grillmair {\it et al.}, AJ {\bf 109}, 2553 (1995)

\bibitem{ibata+01}R. Ibata {\it et al.}, Nature {\bf 412}, 49 (2001)

\bibitem{paper2b}K.V. Johnston, P.I. Choi \& P. Guhathakurta, AJ submitted (2002)

\bibitem{leon+00}S. Leon, G. Meylan \& F. Combes, A\&A {\bf 359}, 907 (2000)

\bibitem{majewski+00}S. Majewski {\it et al.}, AJ {\bf 119}, 760 (2000)

\end{thebibliography}
\end{document}